\documentclass[preprint, 12pt,3p]{elsarticle}

\usepackage{amssymb}
\usepackage{graphicx}
\usepackage{float}
\usepackage{caption}
\usepackage{subcaption}
\usepackage{enumitem}
\usepackage{varwidth}
\usepackage{lineno}
\usepackage{fleqn}
\usepackage{amsmath}
\usepackage[normalem]{ulem}
\useunder{\uline}{\ul}{}
\usepackage{threeparttable}
\usepackage{hyperref}
\usepackage{url}

\journal{arXiv}

\begin{document}

\begin{frontmatter}



\title{Social Vulnerabilities and Wildfire Evacuations: A Case Study of the 2019 Kincade Fire}


\author[UF]{Yuran Sun}
\author[UF]{Ana Forrister}
\author[RMIT]{Erica D. Kuligowski}
\author[MU]{Ruggiero Lovreglio\corref{cor1}}
\author[UOU]{Thomas J. Cova}
\author[UF]{Xilei Zhao}

\cortext[cor1]{Corresponding author. School of Built Environment, Massey University, Auckland, 0632, New Zealand. Email: r.lovreglio@massey.ac.nz.}

\address[UF]{Department of Civil and Coastal Engineering, University of Florida, Gainesville, FL 32611, US}
\address[RMIT]{School of Engineering, RMIT University, Melbourne, VIC 3000, Australia}
\address[MU]{School of Built Environment, Massey University, Auckland, 0632, New Zealand}
\address[UOU]{Department of Geography, University of Utah, Salt Lake City, UT 84112, US}

\begin{abstract}
Vulnerable populations are disproportionately impacted by natural hazards like wildfires. It is crucial to develop equitable and effective evacuation strategies to meet their unique needs. While existing studies offer valuable insights, we need to improve our understanding of how vulnerabilities affect wildfire evacuation decision-making, as well as how this varies spatially. The goal of this study is to conduct an in-depth analysis of the impacts of social vulnerabilities on aggregated evacuation decisions, including evacuation rates, delay in departure time, and evacuation destination distances by leveraging large-scale GPS data generated by mobile devices. Specifically, we inferred evacuation decisions at the level of the census block group, a geographic unit defined by the U.S. Census, utilizing GPS data. We then employed ordinary least squares and geographically weighted regression models to investigate the impacts of social vulnerabilities on evacuation decisions. We also used Moran's I to test if these impacts were consistent across different block groups. The 2019 Kincade Fire in Sonoma County, California, was used as the case study. The impacts of social vulnerabilities on evacuation rates show significant spatial variations across block groups, whereas their effects on the other two decision types do not. Additionally, unemployment, a factor under-explored in previous studies, was found to negatively impact both the delay in departure time and destination distances of evacuees at the aggregate level. Furthermore, upon comparing the significant factors across different models, we observed that some of the vulnerabilities influencing evacuation rates for all residents differed from those affecting the delay in departure time and destination distances, which only applied to evacuees. These new insights can guide emergency managers and transportation planners to enhance equitable wildfire evacuation planning and operations. 
\end{abstract}




\end{frontmatter}


\section{Introduction}
\label{sec:intro}
The increasing scale and intensity of wildfires \citep{mccaffrey2018should, liu2010trends, kuligowski2022modeling, bowman2020vegetation} pose an ever-growing threat to the safety, well-being, and health of populations, especially vulnerable communities residing in the wildland-urban interface (WUI) \citep{radeloff2018rapid, reid2016critical,finlay2012health, wong2020can}. In the context of wildfires, vulnerable populations are those more likely to face negative consequences, having limited capacity to prepare for, respond to, and recover from such events \citep{cutter2009social}. Given the escalating risk to these populations, the development of proactive and equitable wildfire evacuation planning strategies is imperative. Nonetheless, the current emergency planning remains incomplete. Taking California as an example, out of 41 counties where emergency planning documents were reviewed, only 8 had free-standing emergency evacuation plans that qualify for evaluation \citep{kano2011local}.

Understanding the impacts of vulnerabilities on evacuation decisions is crucial towards improving emergency planning. Among different types of vulnerabilities \citep{cova2013mapping}, social vulnerability is highlighted as a key factor, with multiple studies emphasizing its significant contribution to evacuation decisions. Social vulnerability can be viewed as a collection of socio-demographic factors such as poverty, disability, elderly, and racial and ethnic minority status that can heighten the potential harm and increase evacuation obstacles to local populations during wildfires. The evacuation decisions include the initial decision of whether to evacuate and subsequent decisions for evacuees such as the time of departure and destination distances \citep{wong2020compliance}.

Among studies examining the impacts of social vulnerabilities on evacuation decisions, the majority focus on the initial choice of whether to evacuate or not \citep{toledo2018analysis, mclennan2019should, mccaffrey2018should}. Factors like being male, having a lower household income, having a lower level of education, being older in age, having a disability, and the presence of elderly individuals in a household have been associated with a decreased likelihood of choosing to evacuate \citep{toledo2018analysis, alsnih2005understanding, wu2022wildfire, mclennan2019should, paveglio2014understanding, kuligowski2020modelling, whittaker2016gendered, katzilieris2022evacuation}. Conversely, studies focusing on how social vulnerabilities affect decisions for evacuees like departure timing and destination distances are relatively scarce. Understanding these aspects is equally crucial for efficient and equitable evacuation planning, as well as for optimizing resource allocation \citep{grajdura2021awareness}. Noteworthy examples include Grajdura et al.'s study on departure timing \citep{grajdura2021awareness} and Wong et al.'s research on destination distances \citep{wong2023understanding}. They found that being male was associated with longer preparation times for evacuation, while higher education levels were linked to choosing more distant evacuation destinations.

In addition to the limited research on the impacts of social vulnerabilities on evacuees' evacuation decisions, another major limitation is the oversight of several important social vulnerabilities such as unemployment and lack of health insurance in existing studies. However, research in the context of other disasters has shown that some of these vulnerabilities can significantly affect behavioral patterns related to evacuation decisions \citep{hong2020modeling}. In studies of hurricanes, it has been shown that the impacts of social vulnerability on mobility patterns during the evacuation period vary spatially \citep{roy2022effect}. Nevertheless, in the context of wildfires, the question of whether different social vulnerabilities have spatial variations in their impact on evacuation decisions remains under-explored. Finally, the majority of research focusing on wildfire scenarios primarily investigates the impacts of social vulnerabilities on specific evacuation decisions. There is a notable lack of research comparing the distinctions and similarities between vulnerabilities that affect the choice to evacuate and those influencing evacuee decisions such as departure timing and destination distances. Conducting such analysis is crucial for developing a more holistic understanding of how social vulnerabilities impact evacuation decisions at various stages. 

This study aims to bridge the aforementioned research gaps and limitations by utilizing the new-generation large-scale mobile device data (GPS trajectories over time). Given the anonymity of the GPS data, which prevents the identification of specific individuals' sociodemographic information, this study focuses on exploring the impacts of social vulnerabilities on evacuation decisions at an aggregate level, along with their spatial variations. The evacuation decisions include evacuation rates, median delay in departure time and long destination distance rates, which refer to the percentages of evacuees whose destinations exceed 50 miles. The study addresses the following research questions: (\textbf{1}) Are social vulnerabilities associated with aggregate-level wildfire evacuation decisions? (\textbf{2}) If yes, to what extent are social vulnerabilities associated with these decisions? (\textbf{3}) Do social vulnerabilities' impacts on aggregate-level evacuation decisions show spatial heterogeneity and how do the coefficient estimates change across different spatial areas?

To answer the three research questions above, this paper uses the 2019 Kincade Fire in Sonoma County, California, as a case study. This study employs a methodology framework similar to that of Wu et al.'s work \citep{wu2022wildfire}, which utilizes GPS and census data to understand aggregate trends at the census block group level, a geographical unit defined by the U.S. Census. However, this study places greater emphasis on examining the impacts of social vulnerabilities on multiple evacuation decisions and their spatial variations. First, the study applies the proxy home location and evacuation behavior inference algorithms developed by \cite{zhao2022estimating} to compute the aggregate-level evacuation rate, median delay in departure time, and long destination distance rate for each census block group. Subsequently, a comprehensive selection of place-based social vulnerabilities was made, guided by Cutter et al.'s \citep{cutter2003social} framework. Additionally, key auxiliary variables were chosen in accordance with the Protective Action Decision Model (PADM) \citep{lindell2012protective} and insights from other existing literature \citep{wu2022wildfire}. These factors were then computed for each census block group within the evacuation zone. Lastly, the study utilized ordinary least squares (OLS) models and conducted Moran's I tests \citep{moran1950notes} on the model residuals. This analysis aimed to assess the influence of social vulnerabilities on evacuation rates, median delay in departure time, and long destination distance rates, while also examining their spatial variations. If a spatial variation was identified, the geographically weighted regression (GWR) model was applied to capture and analyze this variation.

The structure of the paper is as follows: Section 2 offers a literature review that explores the concept of social vulnerabilities, and the relationship between social vulnerabilities and evacuation decisions. It also reviews existing studies that have employed spatial models to capture the spatial variations of contributing factors. Section 3 presents the materials and methods for analyzing the impacts of social vulnerabilities and their spatial variations. Section 4 introduces the selected study sites, provides details on the data used in the analysis, and presents a comprehensive descriptive analysis of the variables. The results and analysis are listed in Section 5. Section 6 presents the findings of the models, compares the results, and formulates policy recommendations based on these comparisons. It also discusses the study's limitations and outlines potential future research directions. Lastly, the conclusion section summarizes the research and its primary findings, and discusses the potential contributions of these findings to the development of evacuation strategies.

\section{Literature Review}
\label{sec::review}

\subsection{Understanding Social Vulnerability}
\label{subsec::social}
The characteristics of a person or community that affect their ability to anticipate, confront, repair, and recover from a natural hazard or human-caused disaster are collectively referred to as social vulnerabilities \citep{flanagan2018measuring}. The social vulnerability paradigm describes the basic assumption that vulnerability reduction is a public good that has been disproportionately provided, as increased vulnerability results from the combination of exposure to hazards and the lack of access to services and resources that inhibit the capacity to cope and recover from the disaster \citep{juntunen2004addressing}. \cite{morrow1999identifying} asserts that emergency planners should identify areas of high risk for vulnerability during disasters to involve them in the planning and response process, and, drawing upon the Hurricane Andrew disaster, highlights that low-income, age, female-headed households, and shorter length of residence are factors that constituted higher risk. In Cutter et al.'s work \citep{cutter2003social}, for the development of the social vulnerability index (SoVI), dimensions of social vulnerability were condensed into 11 underlying factors that were indicative of vulnerability within communities across the United States at the county level; these factors included personal wealth, age, density of the built environment, single-sector economic dependence, housing stock and tenancy, race, ethnicity, occupation, and infrastructure dependence. Others have developed similar indices, \cite{flanagan1792social} developed a social vulnerability index tool for the Centers for Disease Control and Prevention (CDC) and the Agency for Toxic Substances and Disease Registry’s Geospatial Research, Analysis and Services Program. In the study conducted by \cite{flanagan1792social}, social factors attributed to socially vulnerable populations fall into four domains: socioeconomic status, household characteristics and disability, racial and ethnic minority status and language, and housing type and transportation. Socioeconomic status is comprised of income, poverty, employment, and education variables; household characteristics and disability is comprised of age, single parenting, and disability variables; racial and ethnic minority status and language is comprised of race, ethnicity, and English-language proficiency variables; and housing and transportation is comprised of housing structure, crowding, and vehicles access variables \citep{flanagan1792social}.  

\subsection{Social Vulnerability Impacts on Evacuation Decisions}
During a disaster, socially vulnerable populations are of particular concern since they are disproportionately affected by the disaster event due to age, gender (female-identified), socioeconomic status, race or ethnicity, disability, and/or medical condition \citep{wong2020can}. Identification of at-risk populations by emergency managers typically use a self-identification process, a process by which managers partner with community stakeholders, or use social vulnerability tools (i.e., US Census data and Geographic Information Systems), but most have limited training or experience with these tools \citep{wolkin2015reducing}. Among the studies focusing on vulnerability to wildfire events, some notable ones have aimed to identify the key characteristics that make populations vulnerable to wildfires, or determine which characteristics are most influential in the spatial distribution of social vulnerability by GIS mapping of census block groups for a particular region \citep{paveglio2018assessing, palaiologou2019social, wigtil2016places}. Other representative studies have examined key demographic variables that can be considered as potential vulnerabilities, even if they did not explicitly label them as 'social vulnerabilities' \citep{toledo2018analysis, mclennan2019should, mccaffrey2018should, grajdura2021awareness}. 

Although there are some significant works with key findings, the studies examining relationships between social vulnerabilities and subsequent decisions like delay in departure time and destination distances are scarce. \cite{grajdura2021awareness} discovered that gender(male) is associated with longer preparation times for evacuation, while other factors contributing to the timing of departure and preparation are not necessarily related to social vulnerability. \cite{wong2023understanding} found that higher education was the only significant factor contributing to the tendency of individuals to evacuate over longer distances. This correlation is likely due to higher income levels or broader connections with external areas associated with higher educational attainment \citep{wong2023understanding}. Targeted research on social vulnerabilities can provide more valuable insights into whether other factors related to vulnerability have impacts on the timing of departure and the distance to the evacuation destination. 

Several social vulnerability factors, while under-explored in wildfire research, have shown significant effects in studies of other disasters. For instance, in hurricane research, \cite{hong2020modeling} demonstrated that unemployment significantly influences evacuation flows. Furthermore, \cite{whytlaw2021changing} highlighted that lack of health insurance can increase individuals' vulnerabilities, such as limited material or financial resources and restricted access to information during evacuation. In the context of wildfires, these factors should also be given equal significance when analyzing the impact of social vulnerabilities on evacuation decisions.

Furthermore, the only existing effort to investigate the spatial variation in the impacts of social vulnerability has been focused on wildland fire risks \citep{poudyal2012locating}. There is a notable absence of studies examining the variation in impacts on evacuation decisions. However, in disaster studies beyond wildfires, the impacts of social vulnerabilities on evacuation patterns have been shown to exhibit spatial variations. \cite{roy2022effect} found the effects of social vulnerability on taxi trip times during hurricane Sandy varied spatially. Yum's research \citep{Yum2023analyses} indicated that the impacts of explanatory variables on human responses to Winter Storm Kai, which spread snow from the Sierra Nevada to the Northern Plains in 2019, varied across different regions.

\section{Materials and Methods}
This paper specifies and estimates ordinary least squares (OLS) and geographically weighted regression (GWR) models using data at the census-block-group level to investigate the impacts of social vulnerabilities on wildfire evacuation decisions and their spatial variations. The dependent variables in these models are the evacuation rate, median delay in departure time, and long destination distance rate. The census block group level is selected for analysis because it is the smallest geographic scale offering household-level data from the US Census Bureau \citep{us_census_bureau_geography}, which is crucial for accessing social vulnerability information.

The overall study framework, as depicted in Figure 1, comprises four sequential steps. First, we inferred the proxy home locations and evacuation behaviors of mobile device users, and estimated the departure time, delay in departure time, and maximum destination distances of inferred evacuees using the GPS data. Next, we computed the evacuation rates, median delay in departure times, and long destination distance rates at the census-block-group level from the inferred information. We then calculated variables related to social vulnerabilities and key auxiliary variables for each census block group. Finally, we applied three ordinary least squares (OLS) models to investigate the impacts of social vulnerabilities on evacuation decisions at the census block group level. These decisions encompass choices related to whether to evacuate, as well as decision parameters for evacuees, including the delay in departure time and destination distances. Additionally, our analysis examined the spatial variations in these impacts by applying Moran's I test to the residuals of the OLS models. Where spatial variation was identified, it was subsequently captured using the geographically weighted regression (GWR) model.

\begin{figure}[ht]
  \centering
\includegraphics[width=1.0\textwidth]{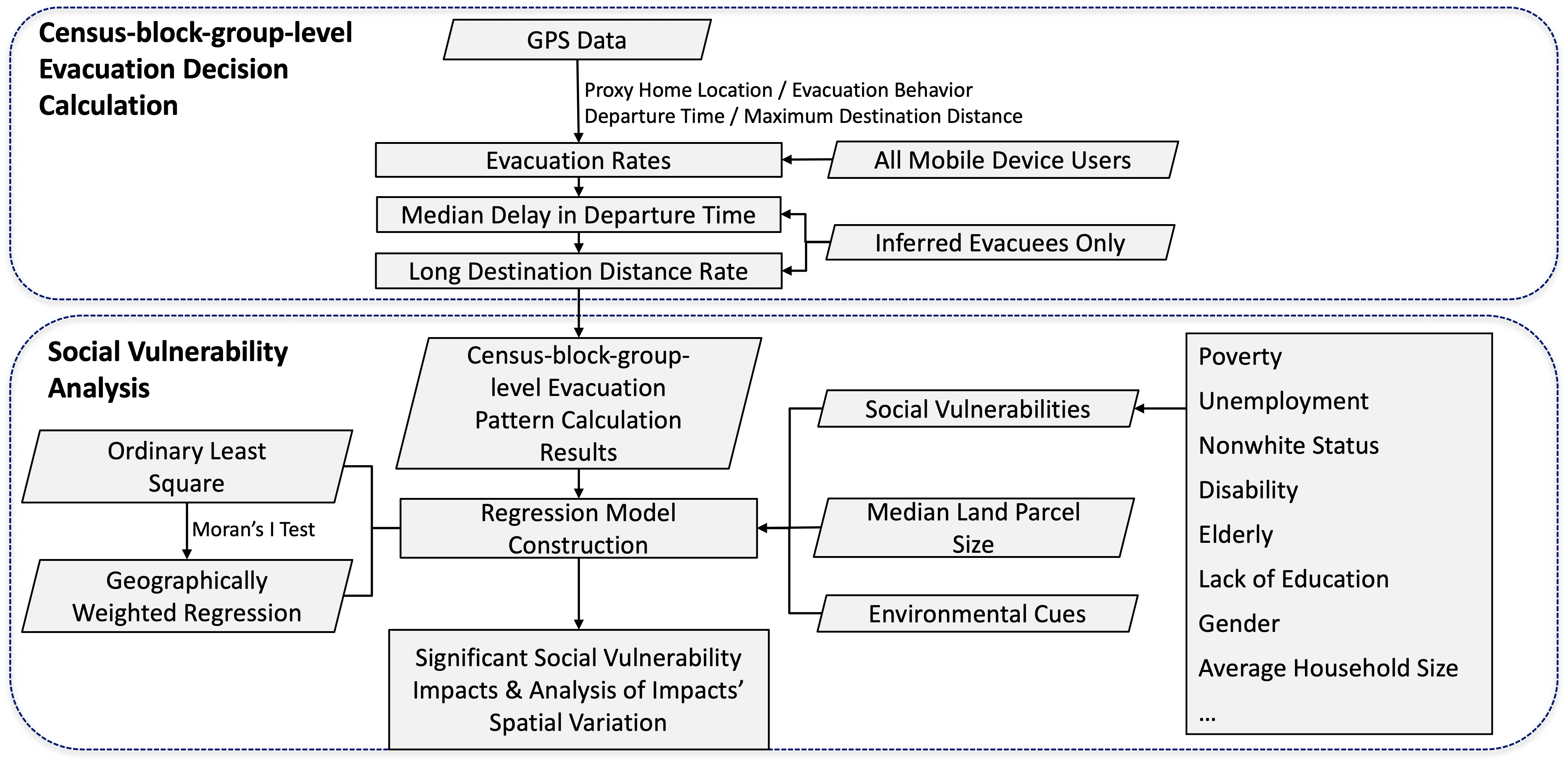}
  \caption{\small Method Overview}
  \label{fig:flow}
\end{figure}

\subsection{Census-block-group-level Evacuation Decision Calculation}
It is important to mention that, due to the anonymized nature of GPS data, the social vulnerability analysis can only be conducted at the aggregate level. This limitation arises from the inability to ascertain the specific sociodemographic characteristics of individual mobile device users. In this manner, the behavior of these users was assumed to be representative at the census block group level, enabling us to conduct aggregate-level analysis. In other words, the evacuation behaviors, delay in departure times, and destination distances of individual mobile device users were consolidated and analyzed at the census block group level. The census block groups selected for social vulnerability analysis are those that intersect with the evacuation zone where evacuation orders or warnings were issued. 

To derive the census-block-group-level evacuation decisions, we first utilized the algorithms proposed by \cite{zhao2022estimating} to infer individual proxy home locations and evacuation behaviors. Specifically, the proxy home location was determined as the centroid of the most frequently occupied 20 × 20 meter cell during the week preceding the wildfire at nighttime. To infer whether a user in the evacuation zone was an evacuee, we observed if he/she left the zone before the lifting of evacuation orders/warnings and subsequently returned to the zone after the orders/warnings were lifted.

The departure time for an evacuee was determined by the first timestamp recorded during their evacuation trip when they exited the 400-meter radius surrounding their home. The 400-meter buffer was established to minimize errors due to data accuracy issues highlighted by \cite{Tom2023}. The delay in departure time \citep{grajdura2021awareness} for evacuees was determined by calculating the difference between the departure time and the moment when the official evacuation orders or warnings were issued in their respective zones. Specifically, the delay in departure time for an evacuee $k$ was defined as:
\begin{linenomath}
  \begin{equation}
  t_k = 
  \begin{cases}
    0 & \text{if } t_{k}^{depart} \leq t_{k}^{awaren} \\
    t_{k}^{depart} - t_{k}^{aware} & \text{otherwise }
\end{cases},
  \end{equation}
\end{linenomath}
where $t_{k}^{depart}$ represents the departure time of the evacuee k, while $t_{k}^{aware}$ denotes the time when the evacuation order/warning was issued for the zone in which the evacuee $k$ was located.

According to \cite{zhao2022estimating}, for each inferred evacuee, the centroid of the 20 × 20 meter cell that was most frequently occupied during the evacuation period each night was considered as one of his/her nighttime stops. Then, the maximum evacuation destination distance was defined as the greatest distance between any of the evacuee's nighttime stops and his/her proxy home location.

Based on the individual-level inference results, we computed the block-group-level evacuation decisions. We derived the definition of the evacuation rate from the work by \cite{wu2022wildfire}. Specifically, for a census block group, the evacuation rate is the percentage of the population within that block group inferred to have evacuated during the wildfire. To calculate the evacuation rates, we first aligned the proxy home location of each user in the study area to their respective block group $i$. The allocation was done through a spatial join between the proxy home locations and the geographical boundaries of the block groups. The evacuation rate for block group $i$ was defined as:
\begin{linenomath}
  \begin{equation}
  R_i = \frac{M_i}{N_i},
  \end{equation}
\end{linenomath}
where $M_i$ represents the total number of inferred evacuees in block group $i$ and $N_i$ refers to the total number of mobile device users in block group $i$.

The median delay in departure time, denoted as $T_{i}$, for block group $i$, was determined by calculating the median of the delay times $t$ of all evacuees residing in block group $i$. Using the median, rather than the average, helps mitigate any bias introduced by extreme departure time values from certain evacuees in particular block groups. Analyzing the delay in departure time rather than the actual departure time helps in assessing evacuees' responsiveness to orders/warnings, which is vital for enhancing future evacuation plans and emergency communication strategies.

The long destination distance rate for a census block group is a metric that depends on the proportion of inferred evacuees whose maximum evacuation distance exceeds 50 miles \cite{younes2021effective}. It reflects the capacity and willingness of evacuees within a particular census block group to evacuate to distant areas. Specifically, for the block group $i$, we defined the long destination distance rate as:
\begin{linenomath}
  \begin{equation}
  D_i = \frac{L_i}{N_i},
  \end{equation}
\end{linenomath}
where $L_i$ is the number of inferred evacuees whose maximum destination distance exceeds 50 miles.

\subsection{Model Specification}
After determining the census-block-group-level evacuation rates, median delay in departure times, and long destination distance rates, the subsequent step involved the generation of independent variables pertaining to social vulnerabilities along with other essential variables. To accomplish this, we initially selected the eligible census block groups intersect with the evacuation zone by conducting a spatial join. Then, we integrated the associated independent variables into the dataset based on their corresponding GEOID, which is the unique identifier code for each census block group.

The variables related to social vulnerabilities were chosen based on relevant studies in the literature \citep{cutter2003social, flanagan1792social, yabe2020effects, wu2022wildfire}. To be specific, the variables included age, poverty, employment status, gender, education, household composition, disability, minority status, insurance coverage, housing, and transportation \citep{cutter2003social, flanagan1792social}. Key additional variables were selected based on the Protective Action Decision Model (PADM) introduced by \cite{lindell2012protective} and findings from a relevant study in the current literature \citep{wu2022wildfire}. The PADM model delineates that people's decision to evacuate is influenced by environmental cues. Specifically, we considered the distance from a particular census block group to the wildfire, as this distance affects the extent to which individuals physically perceive the onset of the fire, which is considered an environmental cue. Furthermore, we included the median land parcel size for each block group, which significantly impacts evacuation decisions in \cite{wu2022wildfire}'s work, as an essential built-in environmental factor.

The final dataset was first employed to create three OLS models. These models pertain to the evacuation rate, median delay in departure time, and long destination distance rate. The equations for OLS models are formulated as:
\begin{linenomath}
  \begin{equation}
  R_i = \beta_{R0}+\beta_{R1}X_{Ri1}+\cdots+\beta_{Rp}X_{Rip}+\epsilon_{Ri},
  \end{equation}
\end{linenomath}
\begin{linenomath}
  \begin{equation}
  T_i = \beta_{T0}+\beta_{T1}X_{Ti1}+\cdots+\beta_{Tp}X_{Tip}+\epsilon_{Ti},
  \end{equation}
\end{linenomath}
\begin{linenomath}
  \begin{equation}
  D_i = \beta_{D0}+\beta_{D1}X_{Di1}+\cdots+\beta_{Dp}X_{Dip}+\epsilon_{Di}.
  \end{equation}
\end{linenomath}
In these equations, $R_i$, $T_i$, and $D_i$ are continuous variables representing the evacuation rate, median delay in departure time, and long destination distance rate for census block group $i$ respectively. Both $R_i$ and $D_i$ range from 0 to 1, while $T_i$ was measured in hours. $\beta_{R0}$, $\dots$,$\beta_{Rp}$, $\beta_{T0}$, $\dots$, $\beta_{Tp}$, $\beta_{D0}$, $\dots$, $\beta_{Dp}$ are the parameters to be estimated. Here, \textit{p} denotes the total number of block groups within the evacuation zone. $X_{Rij}$, $X_{Tij}$, $X_{Dij}$ are the $j_{th}$ independent variables of census block group $i$ in the corresponding evacuation rate, median delay in departure time, and long destination distance rate models, respectively. $\epsilon_{Ri}$, $\epsilon_{Ti}$, $\epsilon_{Di}$ are the error terms. 

We then extracted the residuals from the OLS models and conducted the Global Moran's I test, to examine the spatial autocorrelation of the residuals. This test's null hypothesis posits the absence of spatial autocorrelation. The resulting Moran's Index quantifies the extent of spatial clustering in the residuals. In our study, we set the threshold for rejecting the null hypothesis at a significance level of 0.05. The rejection of the null hypothesis indicates the presence of spatial autocorrelations in the OLS model residuals, suggesting that a transition to spatial models would be more appropriate.

GWR models, as spatial models, effectively address the issue of spatial autocorrelation by capturing the inherent spatial heterogeneity in coefficient estimates. Their model structures closely resemble the OLS model, with the added capability to accommodate locally varying coefficient estimates for study variables. The equations of the GWR models are formulated as:
\begin{linenomath}
  \begin{equation}
  R_{i(u_i,v_i)} = \beta_{R0(u_i,v_i)}+\beta_{R1(u_i,v_i)}X_{Ri1}+\cdots+\beta_{Rp(u_i,v_i)}X_{Rip}+\epsilon_{Ri},
  \end{equation}
\end{linenomath}
\begin{linenomath}
  \begin{equation}
  T_{i(u_i,v_i)} = \beta_{T0(u_i,v_i)}+\beta_{T1(u_i,v_i)}X_{Ti1}+\cdots+\beta_{Tp(u_i,v_i)}X_{Tip}+\epsilon_{Ti},
  \end{equation}
\end{linenomath}
\begin{linenomath}
  \begin{equation}
  D_{i(u_i,v_i)} = \beta_{D0(u_i,v_i)}+\beta_{D1(u_i,v_i)}X_{Di1}+\cdots+\beta_{Dp(u_i,v_i)}X_{Dip}+\epsilon_{Di}.
  \end{equation}
\end{linenomath}
($u_i$, $v_i$) denotes the geographic coordinates of the census block group i. $\beta_{R0(u_i,v_i)}$, $\dots$, $\beta_{Rp(u_i,v_i)}$, $\beta_{T0(u_i,v_i)}$, $\cdots$, $\beta_{Tp(u_i,v_i)}$, $\beta_{D0(u_i,v_i)}$, $\dots$, $\beta_{Dp(u_i,v_i)}$ are the parameters at the census block group i to be estimated.

\section{Case Study}
\subsection{Study Site}
The 2019 Kincade Fire in Sonoma County, CA, has been chosen as the case study. Sonoma County is a sizable urban-rural county situated in northern California \citep{countyofsonoma}. According to the U.S. Census Bureau \citep{us_census_bureau_geography}, the estimated population of Sonoma County in 2019 was 494,336, and the median household income was \$87,828. The majority of households fell within the income range of \$50,000 to \$99,999. Sonoma County was one of the two counties in the California Bay area with the highest median age (43.1 years). The white population in Sonoma County made up the majority, accounting for 62.5\% of the total population. Significant disparities in poverty rates across ethnicities existed. Specifically, Hispanics had a significantly higher poverty rate compared to whites or Asians \citep{countyofsonoma}. Given the sociodemographic characteristics outlined above, Sonoma County is a suitable candidate for vulnerability analysis.

\begin{figure}[H]
  \centering
\includegraphics[width=0.6\textwidth]{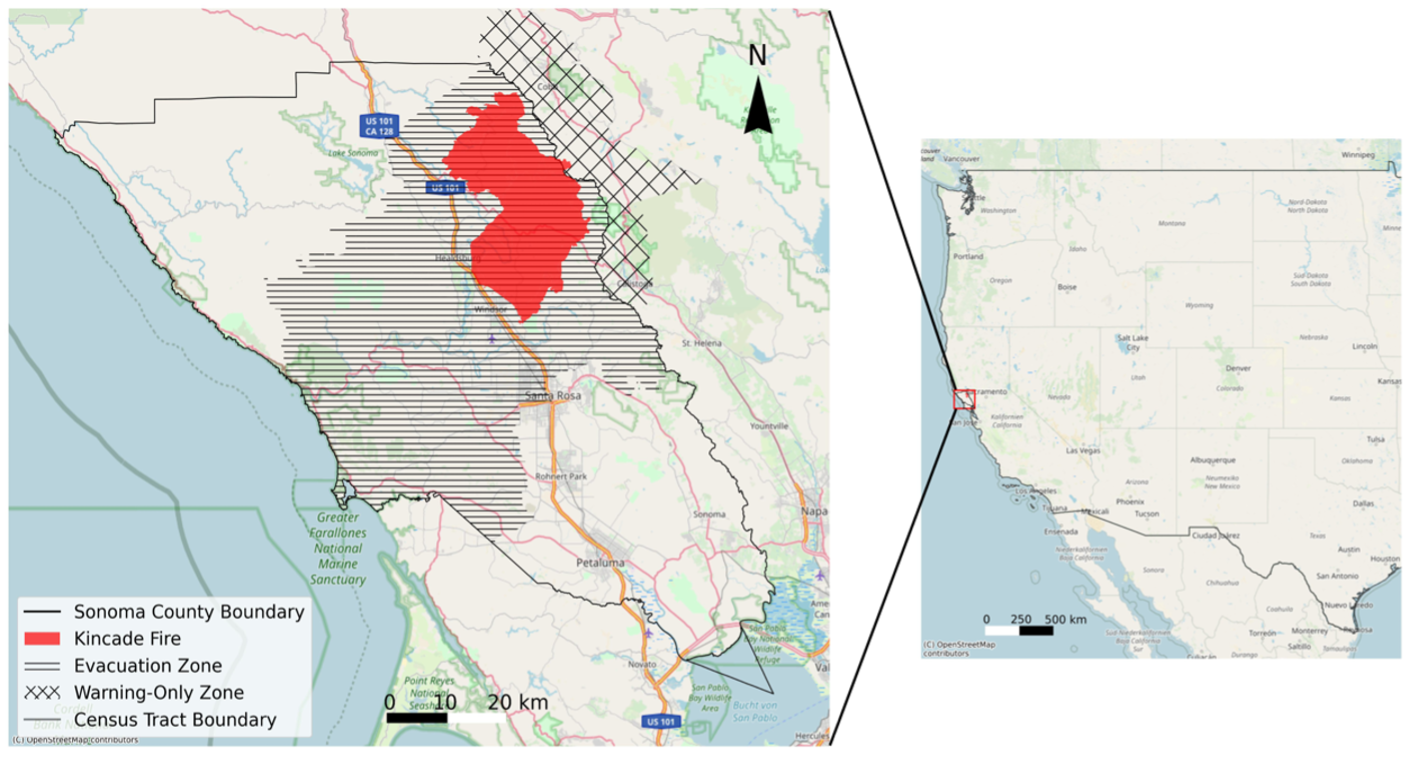}
  \caption{\small Figure of the Study Site}
  \label{fig:site}
\end{figure}

The Kincade Fire originated northeast of Geyserville at 9:27 P.M. on October 23rd, 2019 and was contained by  7:00 P.M. on November 6th, 2019. It was the largest wildfire of the 2019 season in California \citep{sonomacounty}. The fire engulfed an area of 77,758 acres, resulting in damage to 60 structures and complete destruction of 374 structures, and caused injuries to 4 individuals. Throughout the wildfire, more than 186,000 people were prompted to evacuate, marking it as the largest evacuation in Sonoma County \citep{theguardian}. To facilitate the evacuation process, the county was partitioned into zones by the emergency officials. A mandatory evacuation order was first issued in Geyserville on October 26th, followed by a series of orders and warnings for areas stretching to the Pacific Ocean and the northern sections of the City of Santa Rosa \citep{zhao2022estimating}. Figure \ref{fig:site} displays the study site, evacuation areas, and fire parameters.

\subsection{Data}
\begin{figure}[H]
  \centering
\includegraphics[width=0.5\textwidth]{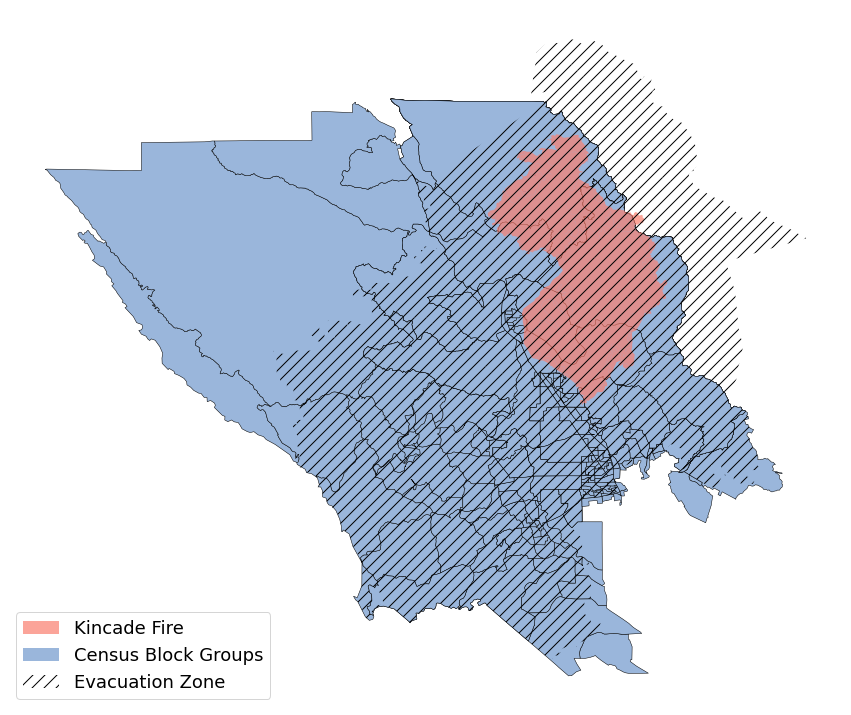}
  \caption{\small Census Block Groups and Kincade Fire Parameter}
  \label{fig:census_block_group}
\end{figure}

The dataset comprises 100,725,936 location records gathered from October 16th, 2019, to November 5th, 2019, in Sonoma County, California. It covers 4736 daily active users (DAUs), each contributing a minimum of 20 records per day. The dataset attributes include unique user IDs, Geohash codes, latitude and longitude coordinates for each data point (ping), and the timestamp (in seconds). Among the 387 census block groups in Sonoma County, a total of 177 block groups are in the evacuation zone (if any portion of the block group lies within the evacuation zone, the entire block group is considered to be within the zone). Out of the 4,736 proxy home locations of the DAUs, 1,475 are situated within block groups in the evacuation zone. Figure \ref{fig:census_block_group} illustrates the census block groups selected in the analysis, as well as the evacuation and the fire areas. 

To address bias stemming from limited sample sizes in certain census block groups, an iterative process was employed to combine them with their nearest neighbors, ensuring that each imputed block group had a minimum of 5 observations. After completing the imputation process, a set comprising 149 imputed census block groups, referred to as Imputed Block Group Set I, was utilized to analyze the relationships between social vulnerabilities and evacuation rates. Among the 2,742 DAUs, only evacuees, a total of 778 individuals, possessed the necessary delay in departure time and maximum destination distance data. Hence, data from these evacuees were applied to compute the aggregate-level inputs for the median delay in departure time and long destination distance rate models. Similarly, to ensure an adequate sample size within each block group, the same imputation process was applied repeatedly until each imputed block group had a minimum of 3 samples. Ultimately, the models analyzing the impacts of social vulnerabilities on median delay in departure times and long destination distance rates included a set consisting of 135 imputed census block groups, referred to as Imputed Block Group Set II.

\subsection{Descriptive Analysis}
Independent variables were sourced and calculated from the U.S. Census Bureau \citep{us_census_bureau_geography}. Table \ref{tab:1} displays the selected variables and their corresponding summary statistics for two imputed block group sets.

\begin{table}[H]
\footnotesize
\caption{Variable Summary Statistics}
\label{tab:1}
\centering
\begin{threeparttable}
\begin{tabular}{lllllllll}
\hline
 & \multicolumn{4}{l}{Imputed Block Group Set I} & \multicolumn{4}{l}{Imputed Block Group Set II} \\ \hline
Variable & Mean & SD & Min & Max & Mean & SD & Min & Max \\ \hline
\multicolumn{9}{l}{\textbf{Dependent Variables}} \\
Evacuation Rate & 0.550 & 0.182 & 0.000 & 1.000 & /  & /  & /  & /  \\
Median Delay in \\ Departure Time (Hours) & / & / & / & / & 2.684  & 3.500  & 0.000  & 17.047  \\
\begin{tabular}[c]{@{}l@{}}Long Destination\\ Distance Rate\end{tabular} & / & / & / & / & 0.396  & 0.226  & 0.000  & 1.000  \\
\multicolumn{9}{l}{\textbf{Independent Variables}} \\
\multicolumn{9}{l}{{\ul Social Vulnerabilities}} \\
Nonwhite Rate & 0.224 & 0.153 & 0.000 & 0.704 & 0.227  & 0.149  & 0.000  & 0.666  \\
Unemployment Rate & 0.047 & 0.046 & 0.000 & 0.270 & 0.043  & 0.039  & 0.000  & 0.167  \\
Poverty Rate & 0.090 & 0.075 & 0.000 & 0.402 & 0.088  & 0.072  & 0.000  & 0.402  \\
Disability Rate & 0.091 & 0.060 & 0.000 & 0.337 & 0.091  & 0.058  & 0.000  & 0.337  \\
Elder Rate & 0.213 & 0.115 & 0.012 & 0.706 & 0.219  & 0.122  & 0.012  & 0.706  \\
\begin{tabular}[c]{@{}l@{}}Bachelor or Above\\ Degree Rate\end{tabular} & 0.354 & 0.150 & 0.054 & 0.659 & 0.349  & 0.147  & 0.054  & 0.659  \\
\begin{tabular}[c]{@{}l@{}}No Health Insurance\\ Rate\end{tabular} & 0.060 & 0.056 & 0.000 & 0.311 & 0.060  & 0.054  & 0.000  & 0.311  \\
Female Rate & 0.510 & 0.064 & 0.223 & 0.696 & 0.511  & 0.061  & 0.223  & 0.683  \\
Children Rate & 0.152 & 0.075 & 0.000 & 0.418 & 0.154  & 0.073  & 0.000  & 0.402  \\
Car Ownership Rate & 0.961 & 0.052 & 0.720 & 1.000 & 0.962  & 0.051  &  0.720 & 1.000  \\
Home Ownership Rate & 0.654 & 0.189 & 0.054 &  1.000 & 0.658  & 0.183  &  0.055 & 1.000  \\
\multicolumn{9}{l}{{\ul Key Auxiliary Variables}} \\
Distance to Fire (km) & 9.804 & 6.674 & 0.000 & 29.718 & 9.427  & 6.588  & 0.000  & 28.477  \\
\begin{tabular}[c]{@{}l@{}}Median Land Parcel \\ Size (Acre)\end{tabular} & 1.011 & 2.367 & 0.018 & 17.965 & 1.024  & 2.436  & 0.024  & 17.965 \\ \hline
\end{tabular}
\begin{tablenotes}
\item Remark: Imputed Block Group Set I is used for analyzing the relationship between social vulnerabilities and evacuation rates, while Imputed Block Group Set II is for studying the impact on median delay in departure times and long destination distance rates
\end{tablenotes}
\end{threeparttable}
\end{table}

The dependent variables consist of Evacuation Rate, Median Delay in Departure Time, and Long Destination Distance Rate, as calculated in accordance with Section 3.2.

For the dependent variables: Across 149 imputed block groups in Imputed Block Group Set I, the evacuation rate varies from 0.000 to 1.000, with an average rate of 0.550. Among the 135 imputed block groups in Imputed Block Group Set II, the maximum median delay in departure time is 17.047 hours, which is less than one day, and the average departure time is 2.684 hours; on average, approximately 40\% of the evacuees reached a destination that was at least 50 mile away.

Variables representing social vulnerabilities are related to age, gender, race, education, household composition, vehicle ownership, home ownership, disability status, employment status, poverty, and health insurance coverage. To account for environmental cues and significant built-in factors related to the wildfire, we also included variables including distance to the fire and median land parcel size. Specifically, the unemployment rate indicates the percentage of the population not in the labor force, while the elder rate represents the percentage of the population aged over 65 in the corresponding census block group.

For the social vulnerability-related variables: In two sets of imputed block groups, the average nonwhite rates are closely aligned, being 0.224 and 0.227, with standard deviations of 0.153 and 0.149 respectively. Furthermore, the highest nonwhite rate observed is 0.704 in Set I and 0.666 in Set II. The average unemployment rates in the two sets are 0.047 and 0.043, respectively. These figures are marginally higher than the nationwide unemployment rate of 0.035, as reported by the U.S. Bureau of Labor Statistics \citep{uslabor}. The average elder rates in the two sets, at 0.213 and 0.227 respectively, are noteworthy for also being higher than the nationwide level, which was approximately 0.160 in 2019 \citep{acl2021profile}. Furthermore, in both sets, the poverty rate varies from a minimum of 0.000 to a maximum of 0.402. The average poverty rates for these sets are 0.090 and 0.088, respectively. In the two sets studied, the average rates for children and disability are 0.152, 0.154 for children, and 0.091, 0.091 for disability, respectively. These rates are slightly lower than the nationwide averages, which stand at approximately 0.220 and 0.131 \citep{cdf2021childpopulation, unh2020disabilityreport}.

\section{Results}
We first constructed three OLS models, each respectively using the evacuation rate, median delay in departure time, and long destination distance rate as the dependent variables. In each model, we evaluated multicollinearity using the Variance Inflation Factor (VIF), with all VIF values found to be less than 5, which indicates that there are no significant concerns regarding multicollinearity in these models.

Table \ref{tab:3} display both the summary statistics for estimated coefficients and the significance levels of each coefficient in the three OLS models.

\begin{table}[!ht]
\centering
\footnotesize
\caption{Ordinary Least Square Model Results}
\label{tab:3}
\begin{threeparttable}
\begin{tabular}{lllllll}
\hline
 & \multicolumn{2}{l}{Evacuation Rate} & \multicolumn{2}{l}{\begin{tabular}[c]{@{}l@{}}Median Delay in \\ Departure Time\end{tabular}} &\multicolumn{2}{l}{\begin{tabular}[c]{@{}l@{}}Long Destination \\ Distance Rate\end{tabular}} \\ \hline
Variable & Coef & Std Err & Coef & Std Err & Coef & Std Err \\ \hline
Constant & 0.611 & 0.369 &-15.618 & 9.344 &0.655 & 0.492  \\
Nonwhite Rate & -0.336* & 0.132 & 2.129 & 3.269 & -0.264 &0.172 \\
Unemployment Rate & 0.436 & 0.341 & 18.123\~ & 9.705 & -0.128* & 0.511 \\
Poverty Rate & -0.422* & 0.211 & 0.620 & 5.341 & -0.582* & 0.281 \\
Disability Rate & -0.305 & 0.270 & 14.865* & 6.908 & 0.047 & 0.364 \\
Elder Rate & -0.016 & 0.196 & 10.950* & 4.822 & 0.554* & 0.254 \\
\begin{tabular}[c]{@{}l@{}}Bachelor or Above\\ Degree Rate\end{tabular} & 0.137 & 0.140 & 3.386 & 3.476 & -0.295 & 0.183 \\
\begin{tabular}[c]{@{}l@{}}No Health Insurance\\ Rate\end{tabular} & 0.360 & 0.299 & -1.947 & 7.443 & 0.405 & 0.392 \\
Female Rate & 0.137 & 0.237 & 1.307 & 6.042 & -0.510 & 0.318 \\
Children Rate & -0.144 & 0.266 & 20.206** & 6.961 & 0.219 & 0.367 \\
Car Ownership Rate & 0.075 & 0.358 & 12.417 & 8.903 & 0.176 & 0.469 \\
Home Ownership Rate & -0.088 & 0.101 & -1.692 & 2.532 & -0.004 & 0.133 \\
Distance to Fire (km) & -0.007*** & 0.002 & -0.184 & 0.060 & -0.009** & 0.003 \\
\begin{tabular}[c]{@{}l@{}}Median Land Parcel \\ Size (Acre)\end{tabular} & 0.002 & 0.006 & -0.074 & 0.153 & 0.002 & 0.008 \\ \hline
\end{tabular}
\begin{tablenotes}
\item *** \textless 0.001 ** \textless 0.01 * \textless 0.05 \textasciitilde \textless 0.1; $N_{EvaRate}$ = 149, $N_{Delay}$=135, $N_{LongDes}$=135, $R^{2}_{EvaRate}$ = 0.19, $R^{2}_{Delay}$=0.22, $R^{2}_{LongDes}$=0.22
\end{tablenotes}
\end{threeparttable}
\end{table}

In the evacuation rate OLS model, two out of the nine independent variables related to social vulnerabilities show statistical significance. Specifically, poverty rate ($\beta=-0.422$) and nonwhite rate ($\beta=-0.336$) negatively contribute to the evacuation rate. This suggests that block groups with higher concentrations of impoverished or nonwhite individuals tend to have lower evacuation rates. Regarding the environmental cues, distance to fire exhibits high significance. Block groups that are farther away from the fire tend to exhibit lower evacuation rates ($\beta=-0.007$).

In the OLS model for median delay in departure time, three social vulnerability factors exhibit significant contributions, while one social vulnerability factor displays marginal significance. unemployment ($\beta = 18.123$), disability ($\beta=14.865$), elder ($\beta=10.950$) and children rate ($\beta=20.206$) all positively influence the median delay in departure time. These results indicate that evacuees in block groups with higher concentrations of unemployed, disabled, elderly, and child populations tend to experience longer departure time delays. For the environmental cue, the distance to the fire does not significantly influence the delay in departure time. This suggests that once evacuees decide to evacuate, the distance to the fire does not have a significant effect on their response time after receiving the order/warning.

In the long destination distance rate OLS model, three variables related to social vulnerability have significant effects. Among them, unemployment ($\beta=-0.128$) and poverty rates ($\beta=-0.582$) negatively contribute to the long destination distance rate, while the elder rate ($\beta=0.554$) has a positive impact on the long destination distance rate. It shows a lower likelihood of distant evacuation among evacuees from block groups with higher rates of poverty and unemployment, whereas those from areas with a larger elderly population tend to evacuate to more distant locations.

Results of the Moran's I test, detailed in Table \ref{tab:moran}, reveal spatial autocorrelation in the residuals of the evacuation rate model, indicated by a Moran's I value of 0.177 and a p-value of 0.001. This suggests a transition to a spatial model for the evacuation rate is necessary. The Moran's I test results for the median delay in departure time and the long destination distance rate yield p-values of 0.130 and 0.142, respectively, both exceeding the 0.05 threshold. It indicates that the hypothesis of no spatial autocorrelation in these variables cannot be rejected. This suggests that the absence of spatially-varying predictors in the model is likely, indicating that Ordinary Least Squares (OLS) regression could be sufficiently effective in capturing the impacts of social vulnerability factors.

\begin{table}[!ht]
\centering
\caption{Moran's I Test Results}
\label{tab:moran}
\begin{threeparttable}
\begin{tabular}{lllll}
\hline
 & OLS &  & \multicolumn{2}{l}{GWR} \\ \hline
 & Moran's I & p value & Moran's I & p value \\ \hline
Evacuation Rate & 0.177 & 0.001 & 0.048 & 0.124 \\
\begin{tabular}[c]{@{}l@{}}Median Delay in\\ Departure Time\end{tabular} & 0.043 & 0.130 & / & / \\
\begin{tabular}[c]{@{}l@{}}Long Destination\\ Distance Rate\end{tabular} & -0.059 & 0.142 & / & / \\ \hline
\end{tabular}
\end{threeparttable}
\end{table}

To capture the spatial variation in how social vulnerability factors impact the evacuation rate, a Geographically Weighted Regression (GWR) model was constructed. Table \ref{tab:6} displays the mean, standard deviation, minimum, and maximum values of the model's coefficient estimates.

\begin{table}[!ht]
\centering
\footnotesize
\caption{Geographically Weighted Regression Model Result}
\label{tab:6}
\begin{threeparttable}
\begin{tabular}{llllll}
\hline
 & \multicolumn{5}{l}{Evacuation Rate} \\ \hline
Variable & Mean & Std & Min & Mean & Max\\ \hline
Constant & 0.321 & 0.591 & -1.213 & 0.332 & 1.469 \\
\textbf{Nonwhite Rate} & \textbf{-0.177} & \textbf{0.431} & \textbf{-0.604} & \textbf{-0.343} & \textbf{0.841} \\
Unemployment Rate & 0.394 & 0.522 & -0.918 & 0.692 & 0.893 \\
\textbf{Poverty Rate} & \textbf{-0.298} & \textbf{0.252} & \textbf{-0.667} & \textbf{-0.376} & \textbf{0.303} \\
Disability Rate & -0.088 & 0.323 & -0.531 & -0.249 & 0.910 \\
Elder Rate & -0.079 & 0.139 & -0.353 & -0.087 & 0.202 \\
\begin{tabular}[c]{@{}l@{}} Bachelor or Above\\ Degree Rate\end{tabular} & 0.099 &  0.215 & -0.188 & 0.029 & 0.696 \\
\begin{tabular}[c]{@{}l@{}}No Health Insurance\\ Rate\end{tabular} & 0.406 & 0.289 & -0.287 & 0.516 & 0.712 \\
Female Rate & 0.229 & 0.248 & -0.223 & 0.188 & 0.881 \\
Children Rate & -0.247 & 0.305 & -1.071 & -0.160 & 0.344 \\
Car Ownership Rate & 0.322 & 0.663 & -1.344 & 0.444 & 1.621 \\
Home Ownership Rate & -0.163 & 0.110 & -0.326 & -0.166 & 0.078 \\
\textbf{Distance to Fire (km)} & \textbf{-0.007} & \textbf{0.013} & \textbf{-0.028} & \textbf{-0.009} & \textbf{0.019} \\
\begin{tabular}[c]{@{}l@{}}Median Land Parcel \\ Size (Acre)\end{tabular} & 0.012 & 0.049 & -0.077 & 0.010 & 0.090 \\ \hline
\end{tabular}
\begin{tablenotes}
\item Remark: Bold font indicates that statistical significance at the 0.05 level has been achieved in at least one imputed census block group.
\end{tablenotes}
\end{threeparttable}
\end{table}

The $R_{EvaRate}^2$ improved from 0.19 in the OLS model to 0.53 in the GWR model. Additionally, the GWR model reduced Moran's I value from 0.177 to 0.048, rendering the Moran's I test on the residuals of the GWR model insignificant. These improvements suggest that the GWR model effectively captures spatial autocorrelations, ensuring no significant spatial-varying predictors are missing. Consequently, the GWR model demonstrates a better fit by accounting for these spatial autocorrelations.

\begin{figure}[ht]
     \centering
     \begin{subfigure}{0.40\textwidth}
         \centering
         \includegraphics[width=1.0\textwidth]{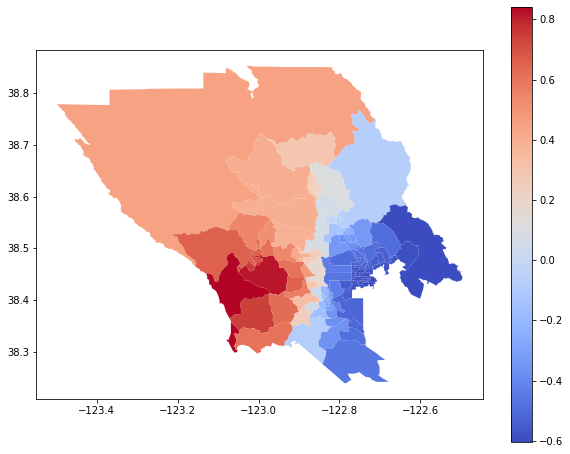}
         \caption{Nonwhite Rate Estimates}
         \label{fig:nonwhite_rate}
     \end{subfigure}
     \begin{subfigure}{0.40\textwidth}
         \centering
         \includegraphics[width=1.0\textwidth]{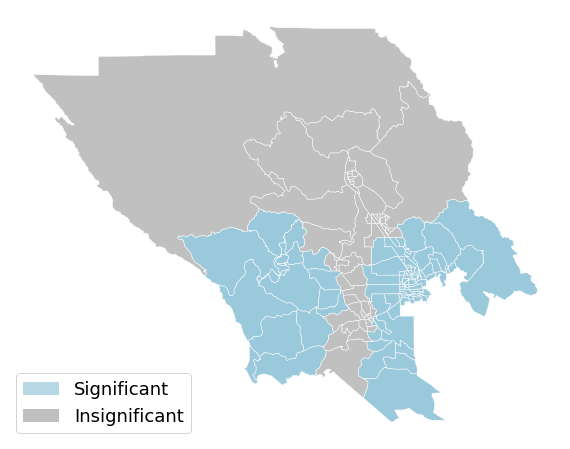}
         \caption{Significance of Nonwhite Rate}
         \label{fig:nonwhite_rate_p}
     \end{subfigure}\\
     \begin{subfigure}{0.40\textwidth}
         \centering
         \includegraphics[width=1.0\textwidth]{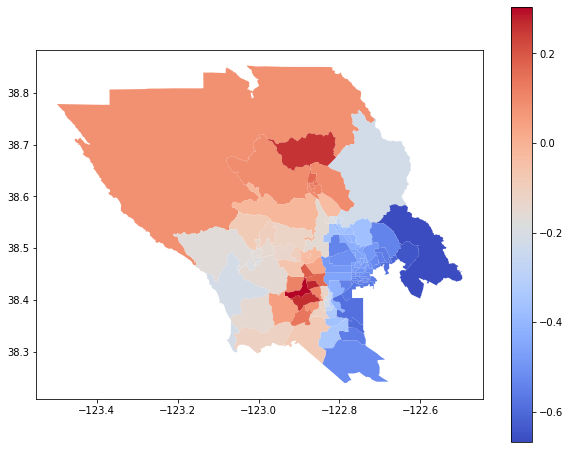}
         \caption{Poverty Rate Estimates}
         \label{fig:poverty_rate}
     \end{subfigure}
     \begin{subfigure}{0.40\textwidth}
         \centering
         \includegraphics[width=1.0\textwidth]{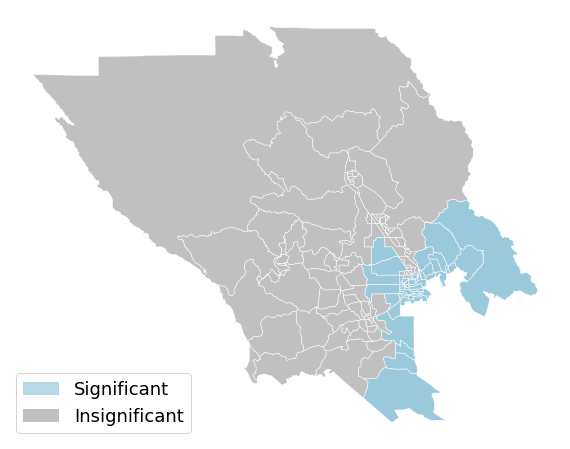}
         \caption{Significance of Poverty Rate}
         \label{fig:poverty_rate_p}
     \end{subfigure}
     \caption{Estimates and Significance of Variables of the Evacuate Rate Model}
     \label{fig:vulnerability_rate}
\end{figure}

 A notable finding from the Geographically Weighted Regression (GWR) results, which contrasts with the Ordinary Least Squares (OLS) model, is the variability in the coefficients for nonwhite and poverty rates across different census block groups. To further explore this, the estimates and significance of these three key variables related to social vulnerability were plotted and are displayed in Figure \ref{fig:vulnerability_rate}. It is important to emphasize that in this analysis, estimates were only considered significant if their associated p-values were less than 0.05.

Observations from Figure \ref{fig:vulnerability_rate} reveal that within the nonwhite rate variable, significant negative estimates are found only in block groups located in the southwest and southeast of the evacuation zone. The results partially align with the OLS model, showing that in the southeast block groups, higher nonwhite rates negatively impact the evacuation rate. Besides, significant negative impacts of poverty rates are observed exclusively in the southeast block groups.

\section{Discussion}

In this work, we analyzed and identified social vulnerability factors that affect aggregate-level evacuation decisions in terms of evacuation rate, median delay in departure time, and long destination distance rate.

The results of the evacuation rate model show that several factors can affect this rate, such as the nonwhite and poverty features of census blocks. This finding aligns with previous research on hazards and disaster vulnerability while providing new findings for the field of wildfire evacuation research. For the block groups located in the southeast of the evacuation zone, the finding that nonwhite populations are less likely to evacuate aligns with the research conducted by \cite{davies2018unequal}, as their study indicates that nonwhite populations are generally less inclined to embrace wildfire migration practices when compared to white communities. Additionally, for the southeast block groups, higher poverty rates contributing to lower evacuation rates is consistent with the finding of \citet{cutter2003social}. It is important to note that the negative impacts of both nonwhite status and poverty are only significant for southeast block groups. This is likely due to weaker social connections and less established disaster preparedness plans specifically among populations with these vulnerabilities, a situation reflected by the higher National Risk Index in this area \citep{fema_nri}. Such factors increase the impacts of these social vulnerabilities on the evacuation decision-making process. The above findings indicate that even within the same socially vulnerable group, there may be spatial variations in their evacuation behavior, and evacuation strategies should also take these spatial variations into account.

The findings on the prediction of median delay in departure time also provide insight into some under-explored aspects of wildfire evacuation research. Block groups with a higher proportion of unemployed individuals tend to have longer evacuation delays. This could be related to the characteristics of seasonal fluctuations in jobs of Sonoma County \citep{Pressdemocrat.com}. Unemployed migrant workers, in particular, may encounter evacuation challenges due to factors such as resource constraints and delayed wildfire-related updates. Even if they are offered official emergency assistance, they may often exhibit reluctance to depend on such assistance, resulting in delayed departures \citep{davies2018unequal}. The observation that the presence of children contributes to increased evacuation delay aligns with the results presented by \citet{hasan2013random, forrister2024analyzing}. This might be attributed to the fact that households with children require additional time to prepare for evacuation, as adults also have the responsibility of caring for the children. The finding that disability is associated with an increased likelihood of delayed departure is consistent with the finding in \citet{golshani2019analysis}. This may be due to the difficulties they experience in following evacuation instructions promptly \citep{ng2015departure}, owing to their mobility restrictions or dependence on assistance from others \citep{golshani2019analysis}. The influence of elderly populations on prolonging the departure time during evacuations has also been discussed in studies by \citet{nakanishi2019investigating}. This could be explained by the fact that elderly people may face physical challenges during evacuation and might psychologically hesitate to evacuate, which could lead to prolonged departure times.

Long destination distance rate is also reflective of the way social vulnerabilities can limit the ability of some to travel further from the threat. Our findings indicate that evacuees from block groups with a higher percentage of impoverished residents tend to evacuate shorter distances. This result aligns with the findings reported by \cite{yabe2020effects}. Further, we identified that a higher unemployment rate correlates with a lower long-destination distance rate, a connection not previously established in the literature. Impoverished and unemployed populations may have fewer resources for evacuation and weaker social and family networks outside their immediate area. These factors can contribute to their tendency to evacuate shorter distances. Finally, elderly populations tend to travel to more distant locations during evacuations. This could be because they prefer to stay with family members who may live farther away rather than opting for nearby accommodations like hotels or friends' homes.

This study underscores both the distinctions and similarities between significant vulnerabilities that influence evacuation rates, and those affecting aggregate-level evacuation decisions only for evacuees (median delay in departure times and long destination distance rates). Specifically, nonwhite status significantly influences the evacuation rates in some block groups and does not significantly affect the median delay in departure time and long destination distance rate. On the other hand, disability and the presence of children show significant contributions only to aggregate-level evacuation decisions for evacuees. These findings indicate that implementing customized approaches to aid diverse vulnerable populations in making different decisions and managing various decision parameters can lead to more effective evacuation strategies. This could involve, for example, offering child-care services and disability-friendly shuttles to ensure the timely evacuation of populations with specific vulnerabilities such as families with young children or individuals with disabilities. 

Additionally, the effects of poverty and nonwhite status on evacuation rates exhibit significant spatial variations. In contrast, the social vulnerability factors influencing aggregate-level decisions for evacuees do not demonstrate such spatial variability. This suggests that the development of flexible evacuation strategies that can be broadly applied to specific vulnerable groups across the entire area, yet are adaptable to meet local needs. 

Furthermore, factors such as poverty, unemployment, and the presence of elderly people significantly influence multiple aggregate-level evacuation decisions. This finding emphasizes the importance for local governments to pay additional attention to these social vulnerability factors when formulating evacuation policies and strategies. Among these factors, unemployment stands out as particularly noteworthy. Despite being relatively understudied in previous literature, it plays a significant role in influencing aggregate-level evacuation decisions for evacuees, including median delay in departure times and long destination distance rates. Recognizing the significance of this factor, it becomes essential to prioritize emergency assistance efforts. This entails providing unemployed workers with timely updates on fire-related situations and essential evacuation resources, and making efforts to accommodate those in need in appropriate shelters. Equally important is building trust in official emergency assistance through regular educational campaigns.

The study's primary limitation stems from using aggregated data, compared to using survey data at the individual or household level. We are assuming that sampled residents of each block group are representative of the area to attribute the socioeconomic and demographic characteristics of the community to the collection of the individual observations of the GPS data. However, this assumption cannot be fully validated, as we do not know any individual-level sociodemographics of the sampled residents due to the nature of the GPS data and privacy concerns. That means that the dataset comprised of the specific users within the study could deviate in their characteristics from the census block group, which could lead to an implicit bias. Therefore, the results of the models are an approximation of which variables are significant predictors of evacuation behaviors. Another limitation arises from the restricted data records, resulting in having to rely on imputed census block groups with only three or five mobile users. This may result in some block groups having too few users, which could make them less representative of the overall population. Additionally, the study relies on the behavioral inference results from the GPS data, which is commonly based on rule-based heuristics and a set of parameters pre-determined by analysts \citep{zhao2022estimating, Tom2023}. This may introduce inaccuracies since the rules and pre-determined parameters may not fully capture the complexity and variability of real-world human behaviors during a wildfire evacuation. The GPS data's reliance on mobile-device users also poses a limitation, as it might not capture the behaviors of non-device users, potentially skewing the representation of the broader population's actions.

%Although the study has the aforementioned limitations, its findings remain valuable for emergency managers and planners. The results can guide the development of targeted policies and strategies to support vulnerable neighborhoods during evacuations. For instance, the study can help identify areas where resources should be allocated to effectively assist vulnerable neighborhoods throughout the evacuation process. Emergency managers could also reach out to vulnerable populations in wildfire areas to see how to better meet their needs.

The use of GPS data allows for more accurate and precise recordings of departure time and destination distance, compared to most survey studies. On the other hand, survey data can vital insights into individual behaviors and can capture the information of non-device users. Therefore, a prospective direction for future studies involves examining the impacts of social vulnerabilities on evacuation decisions by combining survey and GPS data. This approach facilitates a transition from analyzing data at an aggregate level to a more detailed individual level. It also reduces bias associated with neglecting non-device users, a limitation inherent in GPS data. Simultaneously, this method can maintain the spatio-temporal accuracy crucial for evaluating the impact of social vulnerabilities on decisions like departure timing and destination distance. To mitigate the bias resulting from limited data records and behavioral assumptions, future research can also expand on examining how social vulnerabilities influence evacuation decisions across various wildfire scenarios. Additionally, employing Geographically Weighted Regression (GWR) models in future studies across various wildfire scenarios, where spatial autocorrelations are present, is crucial. This approach can improve the model's fitting performance and yield deeper insights into how the impacts of certain factors vary from one block to another, in different study sites with distinct geographic and sociodemographic characteristics.  
\section{Conclusion}
The study utilized GPS data collected during the 2019 Kincade Fire to examine the effects of various social vulnerabilities on evacuation rates, median delay in departure times, and long destination distance rates. By applying ordinary least squares and geographically weighted regression models, we found the significant influence of social vulnerabilities on aggregate-level wildfire evacuation decisions. The results of the evacuation rate model show significant correlations between evacuation rates and social vulnerability factors, including nonwhite and poverty rates, in certain parts of the evacuation zone. Additionally, these results highlight the spatial variations in the impacts of these factors. The median delay in departure time is influenced by unemployment, disability, elder and children rates. Moreover, the long destination distance rate is affected by unemployment, poverty and elder rates. The study further reveals that social vulnerabilities have distinct impacts on evacuation rates and on aggregate-level decisions for evacuees including the median delay in departure time and long destination distance rate. This highlights the necessity for local departments to formulate separate strategies and policies for different evacuation decision processes, ensuring they cater to the specific global and local needs of groups disproportionately affected by certain vulnerabilities. To be specific, for the evacuation decision-making process, efforts could be focused on enhancing evacuation communication and assistance for nonwhite and poor populations, particularly in areas where these factors have been found to negatively impact evacuation rates, to support them in making informed evacuation decisions. For potential evacuees, emergency planners should focus on allocating resources and offering tailored evacuation guidance and services, especially for those who are disabled, elderly, or households with children. Besides, emergency managers could also reach out to these populations to see how to better meet their needs. These approaches ensures evacuees from vulnerable census block groups with high unemployment, disability, elder, poverty and children rates, can evacuate promptly and reach safe and suitable destinations efficiently and effectively.
%\appendix

%\section{Sample Appendix Section}
 %\bibliographystyle{elsarticle-num} 
\bibliographystyle{elsarticle-harv}
\biboptions{semicolon,round,sort,authoryear} 
\bibliography{cas-refs}





% \end{thebibliography}
\end{document}